\documentclass[twocolumn,aps,showpacs,superscriptaddress,longbibliography]{revtex4-1}

\usepackage{bm}
\usepackage{amsmath}
\usepackage{amssymb}
\usepackage[usenames,dvipsnames]{color}
\usepackage{graphicx}
\usepackage{dcolumn}
\usepackage{hyperref}
\usepackage{natbib}
\usepackage[caption=false]{subfig}
\usepackage{float}
\usepackage{times}
\usepackage{todonotes}
\hypersetup{
  colorlinks=true,
  citecolor=blue,
  linkcolor=blue,
  urlcolor=blue}

\begin{document}

\title{Electron correlation effects in enhanced-ionization of diatomic 
       molecules in near-infrared fields}

\author{Siddhartha~Chattopadhyay}
\affiliation{Department of Physics and Astronomy,
             Aarhus University, DK-8000 Aarhus C, Denmark}
\author{Lars~Bojer~Madsen}
\affiliation{Department of Physics and Astronomy,
             Aarhus University, DK-8000 Aarhus C, Denmark}

\date{\today}

\begin{abstract}
We investigate electron correlation effects in internuclear-distance-dependent 
enhanced ionization of $\mathrm{H}_2$, $\mathrm{LiH}$, and 
$\mathrm{HF}$ molecules by intense near-infrared laser pulses using 
a 3D description of the systems with the time-dependent 
generalized-active-space configuration-interaction method. 
This method 
systematically incorporates electron-electron correlation of the quantum 
many-electron system under consideration. Our correlated description of diatomic 
molecules shows that enhanced ionization 
occurs at certain critical internuclear separations and electron correlation 
systematically improves the ionization probability in this process until convergence 
is reached. We demonstrate the failure of the single-active-electron 
and the configuration-interaction singles approximations to produce 
the correct internuclear position and probability of the strong-field enhanced-ionization 
process. We elucidate the role of low-lying electronic excited states in the enhanced 
ionization process of diatomic molecules. There is clear evidence that an 
accurate description of low-lying electronically excited states is important 
to describe the non-perturbative enhanced ionization phenomenon in the ultrashort intense 
near infrared laser pulses.
\end{abstract}
\maketitle

\section{Introduction}
The interaction of atoms and molecules with intense laser fields gives rise to 
ubiquitous phenomena, such as above threshold ionization, 
high-harmonic generation, and enhanced ionization. With progress in  
experimental laser technology, it is now possible to create and observe 
electronic dynamics on their natural time-scale ~\cite{Krausz-RMP-09,
Pazourek-RMP-15}. Along with the experimental progress, theoreticians face the
challenge to accurately describe electron-electron correlation effects 
in strong-field induced dynamics. Current challenges associated 
with the development of time-dependent methods have lead to a series of 
investigations for simple to complex molecular strong-field processes 
with both wave function~\cite{Madsen-18} and density-functional theory 
based methods~\cite{Giovannini-18}.

In the present study, we investigate electron correlation effects in 
enhanced-ionization (EI) in diatomic molecules. EI describes 
the phenomenon that when a molecule is exposed to a strong laser field,  the 
ionization probability increases significantly at certain critical internuclear 
separations. This enhancement is also known as charge-resonance enhanced 
ionization and has been studied extensively both experimentally 
~\cite{Constant-PRL-96,Pavicic-PRL-05,Itzhak-PRA-08,Bocharova-PRL-11,
Wu-NComms-12,Lai-PRAR-14,Xu-NSciRep-15,Erattupuzha-JPB-17} and theoretically
~\cite{Seideman-PRL-95,Zuo-PRAR-95,Chelkowski-JPB-95,Mulyukov-PRA-96,
Plummer-JPB-96,Villeneuve-PRA-96}. The quantum mechanical study of simple 
diatomic molecules with double-well potentials leads to many interesting 
features which are absent in atomic processes. It is well examined that in a 
double-well potential, the electron may localize in one of the potential 
wells with a proper choice of the laser parameters
~\cite{Grossman-PRL-91,Bavli-PRL-92}. This mechanism may also destroy the 
tunneling behavior of the electron between the double wells 
and if  the internuclear separation is  increased,  the localized 
electron may easily tunnel to the continuum from one of the potential wells 
as described in Ref.~\cite{Seideman-PRL-95}. Another mechanism explains EI as 
the strong coupling of charge-resonant states at certain critical internuclear 
separation which then leads to an enhanced molecular ionization probability
~\cite{Zuo-PRAR-95}. Studies by numerically solving the time-dependent 
Schr{\"o}dinger equation (TDSE) show that EI persists in two-electron homo- and 
heteronuclear molecules~\cite{Saenz-PRAR-00,Saenz-PRA-02,Kamta-PRL-05,
Kamta-PRA-07,Dehghanian-PRAR-10,Dehghanian-JCP-13}. An accurate description of 
EI is crucial for the understanding of nuclear kinetic energy release spectra following 
strong-field-induced dissociative ionization (see, e.g., Ref.~\cite{Jing-PRA-16} and references
therein). 

The theoretical research of time-dependent processes in many-electron systems
involves solving the TDSE in the presence of strong laser fields. 
To tackle this problem for more than two electrons, approximations such as, e.g.,
the single-active-electron (SAE)~\cite{Awasthi-PRA-08,
Petretti-PRL-10} and the time-dependent configuration-interaction-singles 
(TD-CIS) approximation~\cite{Rohringer-PRA-06,Greenman-PRA-10,
Karamatskou-PRA-14} are needed. These approximations neglect part of the electron 
correlation effects in the ionization process. The present study on 
diatomic molecules addresses the effects of electron correlation in 
EI using the time-dependent generalized-active-space (TD-GASCI) 
method~\cite{Bauch-PRA-14} in a prolate spheroidal coordinate system
~\cite{Larsson-PRA-16}. Over the past years, various time-dependent 
many-electron methods have been developed to address dynamic electron 
correlation in strong-field ionization of atoms and molecules. Among those, 
the time-dependent $R$-matrix approach
~\cite{VanderHart-PRA-07,Lysaght-PRL-08,Lysaght-PRA-09,VanderHart-PRA-14}, the 
time-dependent Feshbach close-coupling (TDFCC) method~\cite{Sanz-Vicario-PRA-06}, 
the multiconfigurational time-dependent Hartree-Fock (MCTDHF) method
~\cite{Nest-JCP-05,Caillat-PRA-05,Haxton-PRA-11,Haxton-PRA-12,
Hochstuhl-EPJST-14,Greenman-PRA-17}, and the time-dependent restricted-active-space 
self-consistent-field (TD-RAS-SCF) theory~\cite{Miyagi-PRA-13,Sato-PRA-13,
Miyagi-JCP-14,Miyagi-PRA-14,Sato-PRA-15,Miyagi-PRA-17,Omiste-PRA-17,Omiste-PRA-18} 
have been used to understand dynamics. The time-dependent restricted-active-space 
configuration-interaction (TD-RASCI) method~\cite{Hochstuhl-PRA-12}, and the 
time-dependent generalized-active-space configuration-interaction (TD-GASCI) 
method~\cite{Bauch-PRA-14,Chattopadhyay-PRA-15b,Larsson-PRA-16} take electron correlation into 
account through a configuration-interaction (CI) expansion 
by selectively choosing important Slater determinants relevant to the physical 
process of interest. In this method, localized Hartree-Fock and pseudo orbitals are used to 
represent the bound states and grid-based orbitals to obtain an accurate 
description of the continuum states. Depending on the construction of the 
generalized-active-space (GAS) one can reproduce the SAE and CIS approximations
as limiting cases of the TD-GASCI method.

So far the TD-GASCI method has been used to calculate photoelectron spectra, 
ionization yields, structure factors for tunneling ionization, and angle-dependent 
ionization of one- and 
three-dimensional two- and four-electron atoms and molecules~\cite{Bauch-PRA-14,
Chattopadhyay-PRA-15b,Larsson-PRA-16,Yue-PRA-17}. In the present study, we 
employ the method to illustrate electron correlation effects in 
EI of diatomic molecules. First, we consider the simplest possible two-electron 
 molecule, ${\mathrm{H}}_{2}$. This  molecule has been studied 
extensively 
and we use it to check the convergence of the TD-GASCI method by 
comparing with exact TDSE calculations obtained from
Ref.~\cite{Dehghanian-PRAR-10}. To obtain the ionization probability we use linearly 
polarized laser fields with polarization parallel to the the internuclear 
axis within the fixed-nuclei approximation. The role of low-lying electronic 
excited states in EI is studied in detail. Further we consider 
${\mathrm{LiH}}$ and  ${\mathrm{HF}}$ molecules to highlight electron 
correlation effects in EI of multielectron systems. Similarly to the ${\mathrm{H}}_{2}$ case,
we investigate the importance of low-lying electronic excited states in EI. 

The paper is organized as follows. In Sec.~\ref{theory}, we present the TD-GASCI 
method. We elaborate on the construction
of the GAS partitions, define the laser pulses, discuss the calculation of 
the ionization probability and give some remarks on the numerical simulations, Appendix A includes more 
details. In Sec.~\ref{res_dis}, 
we use the TD-GASCI method to elucidate the role of active orbitals in a given 
GAS partition on EI by calculating the ionization probabilities as a 
function of internuclear distance. We consider  different GAS partitions 
which account for electron correlation at different levels of approximation. 
In Sec.~\ref{conc}, we summarize and conclude.

\section{Theory and methodology}
\label{theory}
In this section we briefly present the TD-GASCI method and its 
implementation in prolate spheroidal coordinates, which is discussed in details 
in Refs.~\cite{Bauch-PRA-14,Larsson-PRA-16}. Furthermore the pulses used will be given as well as the form of the complex absorbing potential.

\subsection{TD-GASCI method}

The TDSE for $N_{el}$ -electrons with fixed nuclei reads (we use atomic units throughout)
\begin{equation}
      i\frac{\partial}{\partial t}|\Psi(t)\rangle = H(t)|\Psi(t)\rangle \; .
      \label{tdse}	
\end{equation}
The time-dependent Hamiltonian consists of one- and two-body operators and is 
given by
\begin{equation}
    H(t)=\sum_{i=1}^{N_{el}} {h}_{i}(t)+\sum_{i < j}^{N_{el}} w_{ij} \; ,
    \label{td_ham}
\end{equation}
with the one-body part of electron $i$ given by
\begin{equation}
  {h}_i(t)=-\frac{1}{2}\nabla_i^2 - \frac{Z_1}{|\bm{r}_i - {\bm{R}_{1}}|} - 
            \frac{Z_2}{|\bm{r}_i + {\bm{R}_{2}}|} + \bm{r}_i\cdot \bm{F}(t) \;,
            \label{eq:h1body}
\end{equation}
where $\bm{F}(t)$ is the laser field and $Z_{i} (i=1,2)$ are the charges of 
the two nuclei. In Eq. \eqref{td_ham}, the two-body Coulomb interaction is 
given by
\begin{equation}
 w_{ij} = \frac{1}{|\bm{r}_i - \bm{r}_j|} \;.
\end{equation}
The many-electron wave function is expanded into a basis of time-independent
Slater determinants $|\Phi_I\rangle$,
\begin{equation}
     |\Psi(t)\rangle=\sum_{I \in {\cal V}_{\text{Exc}} }C_I(t) | \Phi_I\rangle 
     \;,
     \label{full_ci_exp}
\end{equation}
where $C_I(t)$ are time-dependent expansion coefficients and $I$ is a 
multi-index, which specifies the configurations from the full Hilbert space 
${\cal V}_{\text{Exc}}$. The Slater determinants are constructed from $N_b$ 
time-independent single-particle spatial orbitals. In terms of 
spin-orbitals we have 2$N_b$ orbitals, $\varphi_i (\bm{r}, \sigma)$, where
a given spatial orbital with different spin-quantum number has the same energy.
After substituting the CI wave function 
from Eq.~\eqref{full_ci_exp} into Eq.~\eqref{tdse}, the TDSE can be expressed 
as
\begin{equation}
     i\frac{\partial}{\partial t} C_I(t) = \sum_{J \in {\cal V}_\text{Exc}}   
	H_{IJ}(t) C_J(t) \;,
    \label{fullci_tdse}
\end{equation}
with the Hamiltonian matrix element, 
$H_{IJ}(t)=\langle \Phi_I |H(t) | \Phi_J \rangle$. 
These matrix elements are constructed by first evaluating 
the one- and two-electron integrals and then rotating  the orbitals as 
described in Ref.~\cite{Bauch-PRA-14}. In the full CI (FCI) method
~\cite{Szabo-MQC-96,Helgaker-MEST-14} one takes into account all possible 
excitations ${\cal V}_{\text{Exc}}$, so that the time-dependent wave function 
reads 
\begin{eqnarray}
   |\Psi_{\rm FCI}(t)\rangle & = & C_0(t)|\Phi_0\rangle + 
	\sum_{ia}C_{i}^{a}(t)|\Phi_i^a\rangle  \nonumber \\
   &  & + \sum_{i<j, a<b}C_{ij}^{ab}(t)|\Phi_{ij}^{ab}\rangle  + \cdots  \;.
   \label{cas-wf}
\end{eqnarray}
Here $i,j,\cdots$ refer to occupied orbitals and $a,b,\cdots$ refer to 
unoccupied orbitals. For example in Eq.~\eqref{cas-wf}, the Slater determinants in the third term, 
$|\Phi_{ij}^{ab}\rangle$, denote doubly excited Slater determinants where 
electrons from orbitals $i, j$ are excited into orbitals $a, b$. 
The FCI expansion is, however, numerically unfeasible even for bound state calculations for 
many-electron systems. In the present case we need to extract the 
ionization probability and it is impractical to treat all electrons with the FCI approach. 
This is due to the exponential scaling in the number of configurations
with the number of basis functions. The GAS concept, 
which was introduced in quantum chemistry~\cite{Olsen-JCP-88}, aims to choose 
the most relevant configurations from the full Hilbert space for the dynamics
under consideration and thus to some extend circumvents the problem of computational 
scaling. In the GAS method, the basis set of Slater determinants is a subset 
of the FCI many-particle basis set, ${\cal V}_\text{Exc}={\cal V}_\text{GAS}$ 
in Eq.~\eqref{full_ci_exp}. By systematically increasing the number of active orbitals,
 we increase the number of Slater determinants, which leads to convergence of the 
method towards the FCI results. This GAS approach not only reduces the 
computational complexity, it also allows an identification
of the most important configurations for a given process and hence helps in
identifying important physics.

\begin{figure}
  \includegraphics[width=\columnwidth]{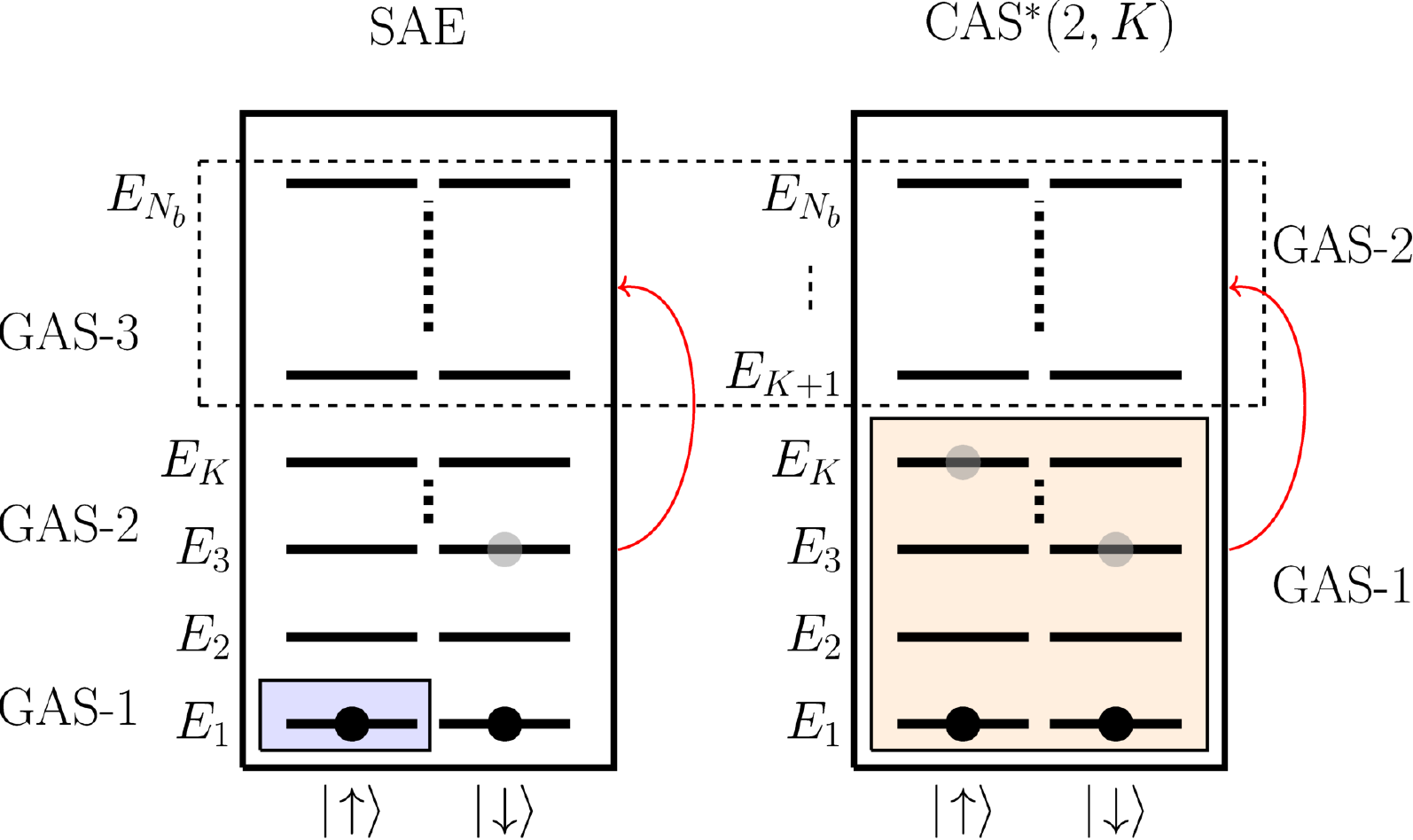}
  \caption{Schematic of GAS partitions used in this work for a two-active 
	   electron molecule. The left panel shows the SAE approximation. 
	   The right panel shows the complete-active space 
	   (CAS$^*$) situation, where all single excitations out of the CAS are included. $E_{i}$ represents the orbital energy of the
	   $i$-th orbital. The SAE approximation consists of three GAS partitions,
	   GAS-1 defines the frozen electrons, GAS-2 defines the single-active-electron, 
	   GAS-3 defines the single excitations from GAS-2.
	   The CAS$^*(2,K)$ notation refers to two-active electrons with $K$ spatial
	   orbitals and the asterisk denotes that all single excitations out of the active space are included. 
	   In the CAS$^*(2,K)$ model we have two GAS partitions. GAS-1
	   contains all possible excitations within this space. GAS-2 defines
	   single-excitation from GAS-1 to GAS-2.}
 \label{fig:gasscheme}
\end{figure}

In Fig.~\ref{fig:gasscheme}, we show the GAS partitions used in this work. 
The energies of the single-particle orbitals are denoted by $E_i$. The two spin
configurations, $|\hspace{-0.1cm}\uparrow\rangle$ and 
$|\hspace{-0.1cm}\downarrow\rangle$, are degenerate in this representation. 
The red arrows imply that only  single ionization is allowed. 
One can obtain the well-known SAE approximation 
from the GAS concept as shown in the left panel of Fig.~\ref{fig:gasscheme}. In 
this illustration, one of the electrons in a two-electron molecule is frozen in the 
GAS-1 space and the other electron is allowed to be excited within the GAS-2 
space. Here we emphasize that only single excitations are allowed in the GAS-2 
space, i.e., for the ionization process we allow only one-electron to be excited from GAS-2 
to the GAS-3 space. The time-dependent wave function in the SAE approximation 
can be written as
\begin{equation}
	|\Psi_{\rm SAE}(t)\rangle = C_0(t)|\Phi_0\rangle + \sum_{a \in vir }
	C_{i}^{a}(t)|\Phi_i^a\rangle \;,
   \label{sae-wf}
\end{equation}
where we note that the sum runs over all virtual orbitals. Here, 
$|\Phi_0\rangle$ is the Hartree-Fock reference determinant, and $|\Phi_i^a\rangle$ is 
a singly-excited determinant. 
Since the sum in Eq.~\eqref{sae-wf} runs over all virtual orbitals, $a$, 
with a fixed core, $i$, it represents an effective interaction felt by the 
single electron, which is created by all the other electrons similar to the 
Hartree-Fock potential. Similarly, the explicit time-dependent wave function 
in the CIS approximation is described within the GAS method as
\begin{equation}
	|\Psi_{\rm CIS}(t)\rangle = C_0(t)|\Phi_0\rangle + \sum_{i \in {\rm occ}}
	\sum_{a \in vir}
	C_{i}^{a}(t)|\Phi_i^a\rangle \;.
   \label{cis-wf}
\end{equation}
Here the sum includes all the core and virtual orbitals and all the 
single-excited Slater determinants are constructed with time-dependent 
coefficient, $C_{i}^{a}(t)$. Note that although we use the same notation for these 
time-dependent coefficient in Eqs.~(\ref{sae-wf}) and (\ref{cis-wf}), they are in 
general different for the different approximation schemes. In the right panel of Fig.~\ref{fig:gasscheme}, 
we show the complete-active-space (CAS) concept~\cite{Olsen-JCP-88,Bauch-PRA-14}, 
which corresponds to a FCI description of the system with a spatial orbital 
index $K$. CAS$^*(2,K)$ refers to two active electrons with $K (2K)$ 
spatial orbitals (spin orbitals) within the given CAS. In this case, all 
possible excitations are treated within the GAS-1 space.  The asterisk denotes that all
single excitations out of the CAS are included. Therefore the 
time-dependent wave function in the GAS method reads
\begin{eqnarray}
	|\Psi_{\rm CAS}(t)\rangle & = & C_0(t)|\Phi_0\rangle + 
	\sum_{i \in occ}\sum_{a \in vir} C_{i}^{a}(t)|\Phi_i^a\rangle  
	\nonumber \\
	&  & + \sum_{i<j \in occ}\sum_{a<b \in vir}
	C_{ij}^{ab}(t)|\Phi_{ij}^{ab}\rangle  + \cdots  \;.
   \label{gas-wf}
\end{eqnarray}

In the frozen-electron approximation, one can freeze the inner core electrons, which
may have an insignificant role on the dynamics. Within the TD-GASCI
method we can create such different models to describe the ionization process
in a many-electron system. In this method, the restriction is 
created on the active space under consideration by choosing $K$ 
spatial orbitals and thus limiting the number of determinants within 
the corresponding GAS partition. We emphasize that the CAS 
notation throughout this  work is accompanied by additional single 
excitations to the final GAS and indicated in our notation by 
CAS$^*(N_{el},K)$, where $N_{el}$ denotes the active electrons and $K$ the 
number of spatial orbitals within the CAS under consideration. 

\begin{figure}
 \includegraphics[width=\columnwidth]{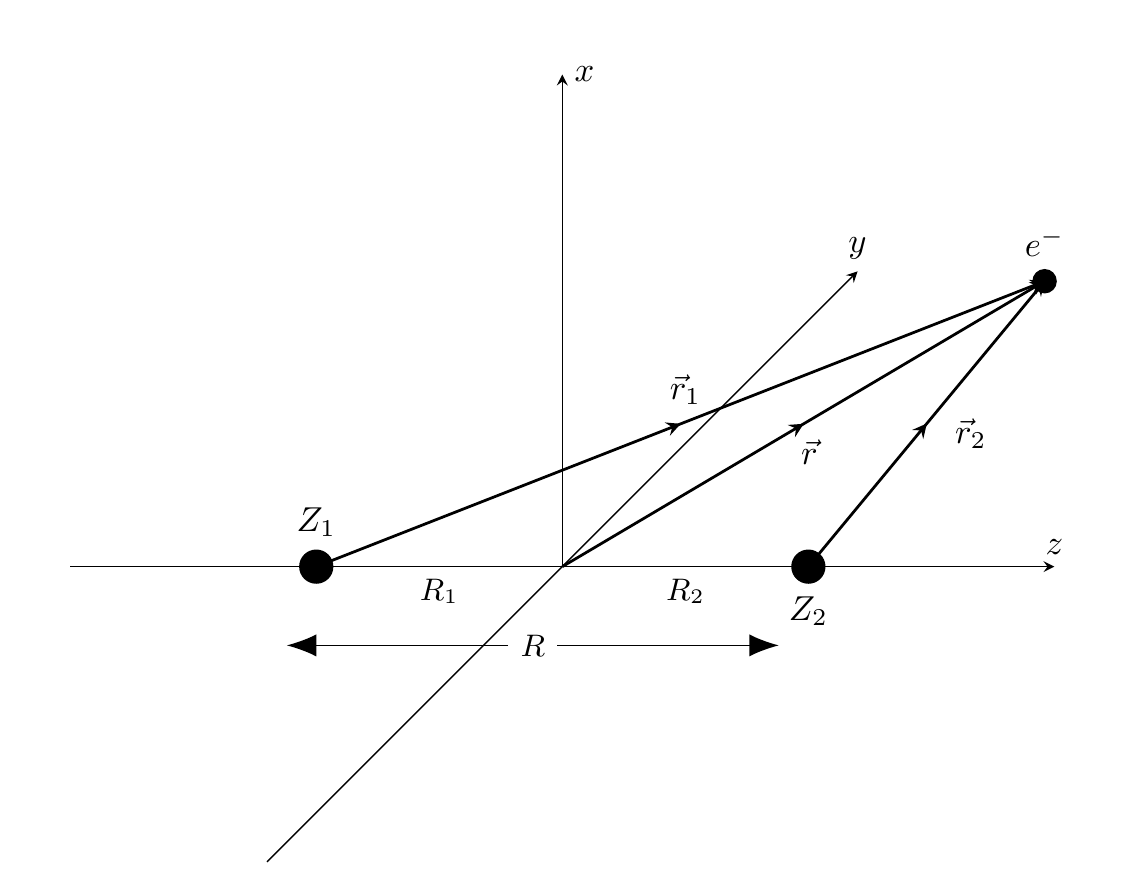}
 \caption{Schematic of the coordinate system for diatomic molecule. 
	  The molecule with binding length $R$ has the origin in the geometric
	  center.} 
 \label{fig:coord-sketch}
\end{figure}

\subsection{Single-particle basis}
The single-particle basis is constructed in the
prolate spheroidal coordinate system. For a detailed
description of the implementation see Refs.~\cite{Larsson-PRA-16,Tao-PRA-09a,Haxton-PRA-11}.
The coordinates are denoted by  $(\xi,\eta,\phi)$, and they are related to 
the Cartesian coordinates by the following relations
\begin{eqnarray}
	\xi  & = & \frac{r_1 + r_2}{R}, \quad \xi  \in (1,\infty) \nonumber \\
	\eta & = & \frac{r_1 - r_2}{R}, \quad \eta \in (-1,1)     \nonumber \\
	\phi & = & \arctan(r_2/r_1),    \quad \phi \in (0,2\pi)   \;.
  \label{eq:prolate-co}
\end{eqnarray}
Here $r_1$ and $r_2$ are the electron coordinates and $R$ is the bondlength
of the diatomic molecule as shown in Fig.~\ref{fig:coord-sketch}. In the 
prolate spheroidal coordinate  system, the time-independent wave function 
is expressed as
\begin{equation}
	\Psi(\xi,\eta,\phi) = \frac{1}{\sqrt{2\pi}}
	\sum_{m} \Psi^{m}(\xi,\eta) e^{im\phi}, m = 0,\pm{1},\pm{2},\cdots \;.
	\label{eq:prolate-wf}
\end{equation}	
Both the $\xi$ and $\eta$-coordinates are described by a finite-element 
discrete-variable-representation (FE-DVR) basis~\cite{Light-AdvChemPhys-07}.
The total simulation box is partitioned into a central and an outer region and 
we use the partially rotated single-particle basis for the whole simulation 
box as discussed in Ref.~\cite{Bauch-PRA-14,Hochstuhl-EPJST-14}. The $\xi$ 
coordinate is partitioned into two regions such that for $\xi < \xi_{s}$ 
the single-particle basis is constructed from localized occupied Hartree-Fock and pseudo
orbitals. For $\xi > \xi_{s}$, FE-DVR 
functions represent the continua.  The domains 
of these coordinates are such that ionization is mainly described by the $\xi$  coordinate, while
the $\eta$ coordinate describes bound-state motion. 

\begin{figure}
 \includegraphics[width=\columnwidth]{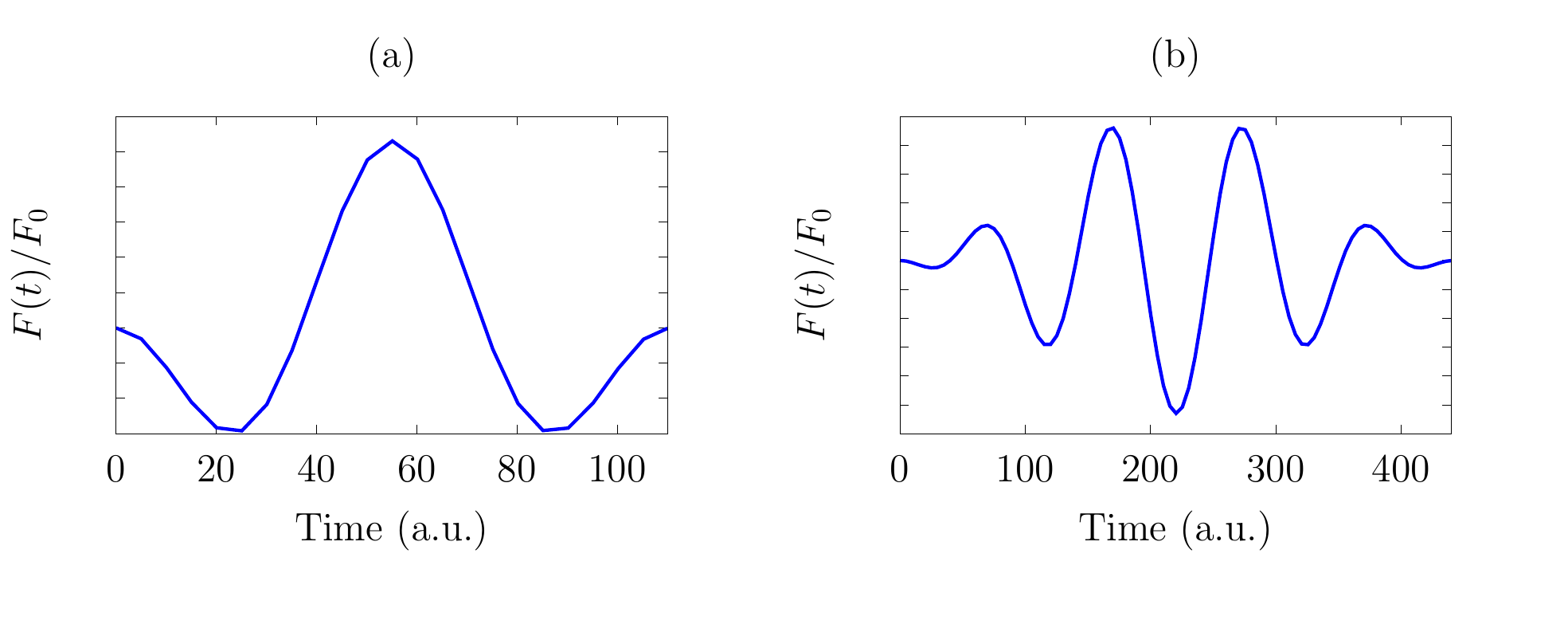}
   \caption{Normalized electric field with (a) a single cycle and 
	    (b) four-cycles.}
 \label{fig:e_field}
\end{figure}

\subsection{Laser pulse parameters}
To study EI, we expose diatomic molecules to 800 nm strong laser
fields, which are described using the length gauge and the dipole approximation. 
The diatomic molecules are 
aligned colinearly with the polarization axis of the laser field. The vector 
potential  has a sine-square envelope
~\cite{Han-PRA-10},
\begin{equation}
   A(t)  = \frac{F_0}{\omega}\sin^2\Big(\frac{\pi t}{T}\Big) 
	   \sin{(\omega t )},
            \qquad (0 \le t \le T) \;,
   \label{vector_pot}
\end{equation}
where $T = N \frac{2\pi}{\omega}$ is the pulse duration with $N$ the number 
of cycles and $\omega$ the angular frequency. The electric field  is obatined as 
$F(t) = - \frac{\partial A(t)} {\partial t}$ and is shown in 
Fig.~\ref{fig:e_field} for the single- ($N = 1$) and four-cycle ($N =4$)
pulses used in the present calculations. $F_0$ is the maximum amplitude of the 
laser pulse.The use of a vector potential to generate the electric field ensures that the time-integral 
over the electric field vanishes once the laser pulse is 
over~\cite{Madsen-PRA-02}.  We note that one needs a pump-probe setup to experimentally 
probe EI in molecules as shown in Ref.~\cite{Xu-NSciRep-15} because the laser 
pulse is so short that a molecule will not have enough time to dissociate to
the critical internuclear distance with EI during the duration of the pulse.

\subsection{Ionization probability}
To extract the total ionization probability, we add a complex absorbing 
potential (CAP) to the full Hamiltonian,
\begin{equation}
       H^{\textup{CAP}}(t)=H(t)-iV_{\textup{CAP}} \;.
\end{equation}
We tested various types of CAPs and found that the following CAP
~\cite{Beck-PhysRep-00} produces a converged ionization probability for all the 
molecules under consideration
\begin{equation}
	V_{\textup{CAP}}(r)= \tilde{\eta} (r-r_{\textup{CAP}})^{b}
	\theta(r-r_{\textup{CAP}}) \;.
      \label{eq:cap}
\end{equation}
Here $\theta$ is the Heaviside step function, which ensures that the CAP is 
switched on once the wave packet reaches $r_{\textup{CAP}}$ and the exponent is set to $b=2$
as in Ref.~\cite{Beck-PhysRep-00}. In 
Eq.~\eqref{eq:cap} $\tilde{\eta}$ is the CAP strength and in the present study
we found converged results with $\tilde{\eta} = 0.5$. Note that for 
the prolate-spheroidal coordinate, we apply the CAP along the $\xi$ coordinate and in all cases $\xi_\text{CAP}=50$. 
The total ionization probability~\cite{Kulander-PRA-87} reads as
\begin{equation}
    \mathcal{P}(t_f)=1-\mathcal{N}(t_f) \;,
    \label{ionization-prob}
\end{equation}
with $\mathcal{N}(t_f) = \langle \Psi(t_f)|\Psi(t_f) \rangle$.
To extract the ionization probability after the end of the pulse, we propagate the 
equations of motion to a final time, $t_f$ = 241 fs. We found that this time is 
sufficient to obtain converged results, also for correlated situations.

\subsection{Remarks on the simulations}
For the numerical simulations, first we prepare the diatomic molecule in its 
ground-state by imaginary-time propagation (ITP). For the ITP, we use 
the short-iterative Arnoldi-Lanczos algorithm~\cite{Beck-PhysRep-00}. Once the 
ground state is converged we apply the laser pulse and propagate in real time. 
We follow the adaptive time-step for the propagation 
of the time-dependent wave packet as discussed in Ref.~\cite{Larsson-PRA-16}. 
The EI  process involves a large number of TD-GASCI simulations for different  
internuclear separations. Therefore, we choose a relatively large inner-region 
of the simulation box such that it retains the converged Hartree-Fock orbitals 
for the time-dependent calculations. For all the diatomic molecules that we 
treated, first we check the convergence of the Hartree-Fock orbitals and 
energies. In the time-dependent simulations we expose the diatomic molecules to 
800 nm ($\omega = 0.057$) laser pulses [Fig.~3]. Both the $\eta$ and $\xi$ coordinates are 
described by FE-DVR functions.  A description of the discretization used for these variables and the 
computational demands is given in Appendix A.

\section{Results and discussion}
\label{res_dis}
In this section we present results on EI for H$_2$, LiF and HF molecules.

\subsection{Two-electron $\mathrm{H}_2$ molecule}
\label{h2}
\begin{figure}
 \includegraphics[width=\columnwidth]{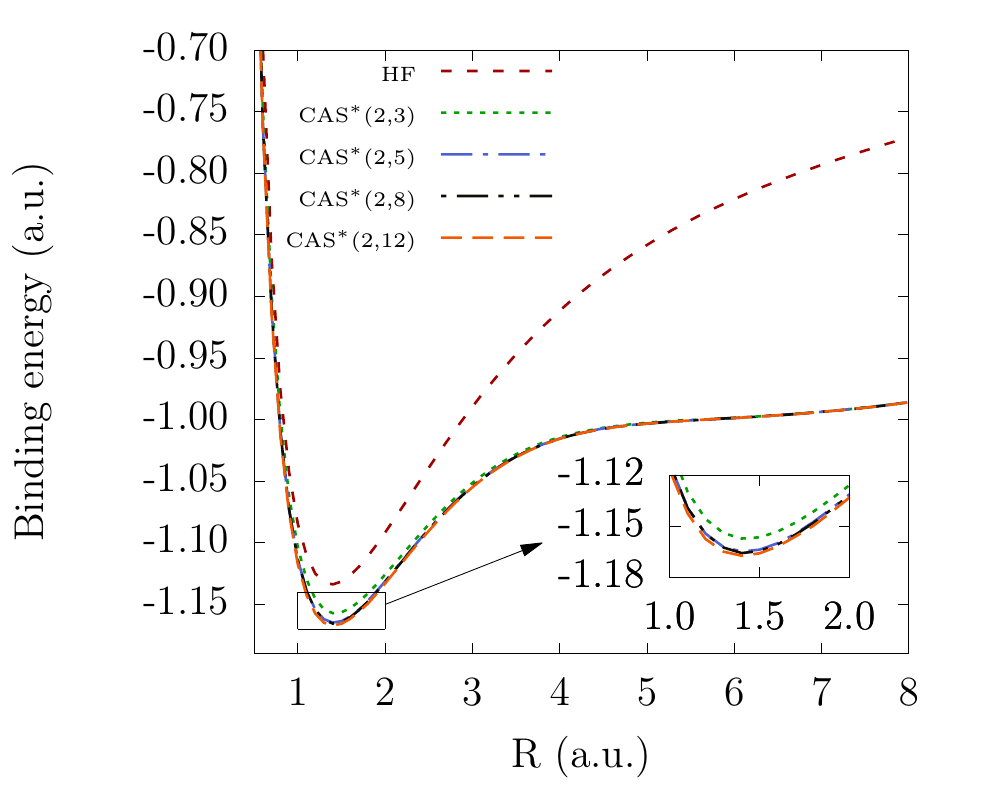}
 \caption{ Ground state energy of $\mathrm{H}_2$ 
	   from imaginary time propagation. CAS$^*(2,K)$, ($K=3,5,8,12$) represents TD-GASCI 
	   with $K$ active spatial orbitals in the CAS. In the figure, the Hartree-Fock result is 
	   denoted by HF. }
 \label{fig:H2-binding}
\end{figure}

In order to study correlation effects in diatomic molecules in connection with EI, the 
two-electron $\mathrm{H}_2$ molecule is a preferred choice as it is 
the simplest system with more than a single electron where the EI has been studied extensively by 
solving numerically the TDSE. First we prepare $\mathrm{H}_2$ in its ground state 
by ITP for a range  of internuclear separations, $R$. In Fig.~\ref{fig:H2-binding} we present the results 
with Hartree-Fock and different GAS partitions. It is seen that 
the CAS$^*(2,3)$-calculations with three active orbitals improve the 
ground-state energy significantly compared to the Hartree-Fock energy. The ground-state energies from SAE and CIS approximations equal the ground-state energy of the Hartree-Fock approach due to  Brillouin’s theorem
~\cite{Szabo-MQC-96}, which states
\begin{equation}
	\langle \Phi_i^a|H_{0}|\Phi_{0} \rangle = 0 \;,
	\label{briltheo}
\end{equation}
with $H_{0}$  the time-independent field-free Hamiltonian.  The CAS$^*(2,5)$ scheme with five active 
orbitals improves the ground-state energy further. To check the convergence of the ground-state 
energy with the number 
of active orbitals, we increase the number of active orbitals in the 
GAS from five to eight and up to twelve for the CAS$^*(2,12)$ model and one 
can see from the figure that the CAS$^*(2,5)$ model is fully converged for 
the ground-state and we also obtain the correct equilibrium bond length
of $R=1.4$ by the ITP method.

\begin{figure}
 \includegraphics[width=\columnwidth]{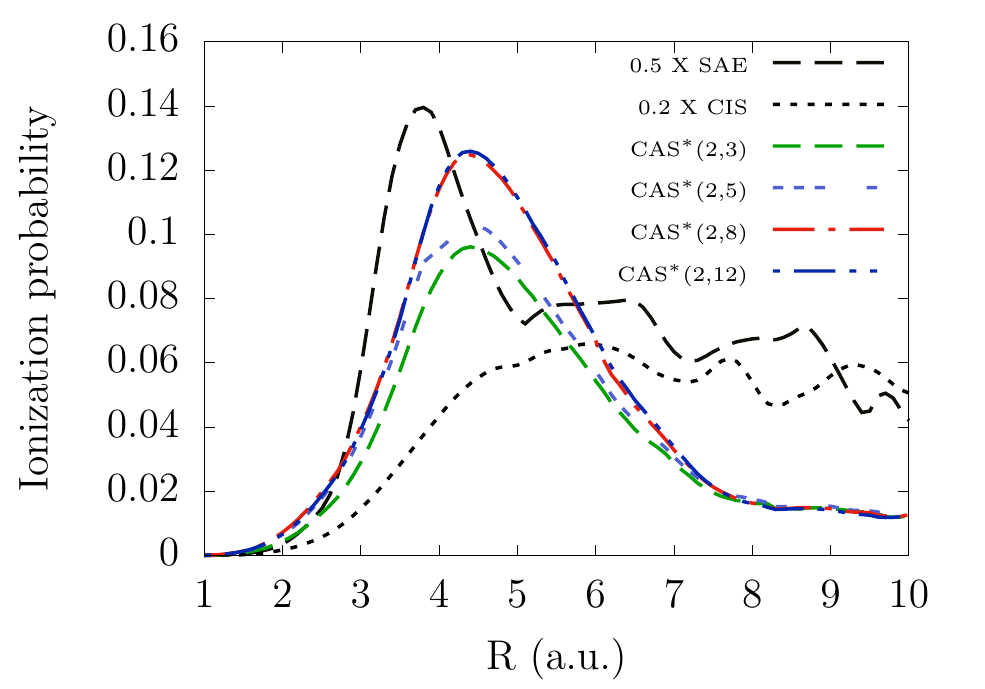}
    \caption{Ionization probability vs internuclear distance $R$ for 
	     $\mathrm{H}_2$ with ${F}_0 = 0.053$ and SAE, CIS and 
	     different GAS approximations. CAS$^*(2,K)$, ($K=3,5,8,12$) represents
	     $K$ active spatial orbitals in the CAS partition. }
    \label{fig:h2-ei}
\end{figure}

For a systematic investigation of the correlation effects in EI of 
$\mathrm{H}_2$, we use the 800 nm single-cycle laser pulse as shown in
Fig.~\ref{fig:e_field} (a) with a peak field strength of ${F}_0 = 0.053$ 
($ 10^{14}$ W/cm$^2$). For an accurate description of the correlation effects 
we take orbitals with higher $m$-quantum numbers as described in 
Eq.~\eqref{eq:prolate-wf} and in the present calculations we have considered 
up to $m = \pm 1$ which produces a converged EI results. In 
Ref.~\cite{Larsson-PRA-16} it was shown that $m = 0, \pm 1$ is sufficient
to obtain a correlated ionization spectra and further increase 
in $m$ does not change the ionization probability significantly. In Fig.~\ref{fig:h2-ei}, we 
present the ionization probability as a function of the internuclear separation.
Here we scaled down the results obtained from the SAE and CIS approximations. 
It is evident that the SAE and CIS approximations do not produce the 
correct EI peak position and magnitude compared to the other correlated 
calculations. Also in these approximations, we observe spurious resonances 
as the internuclear separation is increased from $R =3.8$. 
In all the CAS$^*(2,K)$ models in the figure, the ionization probability increases with the 
increase of internuclear separation and after a critical internuclear 
distance of $R = 4.4$, it decreases and eventually at large internuclear distances equals
the sum of atomic ionization probabilities. The CAS$^*(2,3)$ model is the simplest 
model in the current description of $\mathrm{H}_2$ and it produces 
the EI-peak at the correct position, i.e, $R = 4.4$. As we increase the number 
of active orbitals, the probability converges. Most 
of the correlation contributions are captured in the CAS$^*(2,8)$ model. 
An earlier TDSE calculation produces the EI peak at a similar internuclear distance
~\cite{Dehghanian-PRAR-10}. The same set of laser parameters produced the converged EI 
peak at $R = 4.7$ in our previous 1D calculations ~\cite{Chattopadhyay-PRA-15b}. 
A difference between the 1D and the present calculations is the way the 
Coulomb interaction between the electrons is treated. The regularized Coulomb 
potential in the 1D calculation may overestimate the correlation compared to  
the exact Coulomb interaction. 
We note that the SAE and CIS approximations are inaccurate in describing EI 
both in terms of magnitude and peak position. Similar conclusions were obtained from
1D calculations~\cite{Chattopadhyay-PRA-15b}. As mentioned earlier, the TD-GASCI 
method systematically incorporates the electron correlation in a given 
GAS partition. The main difference  between the SAE and CIS,  and
the GAS calculations is the inclusion of the doubly excited Slater determinants
in the latter case. The SAE and CIS approximations are unable 
to describe EI because they do not include effects of double excitations 
in the many-electron wave function, i.e, the doubly-excited
determinants ($|\Phi_{ij}^{ab} \rangle$) contribute significantly in the
dynamic electron correlation which further underlines the  need for correlated 
many-electron calculations in modeling molecular strong-field ionization processes. 

\begin{figure}
 \includegraphics[width=\columnwidth]{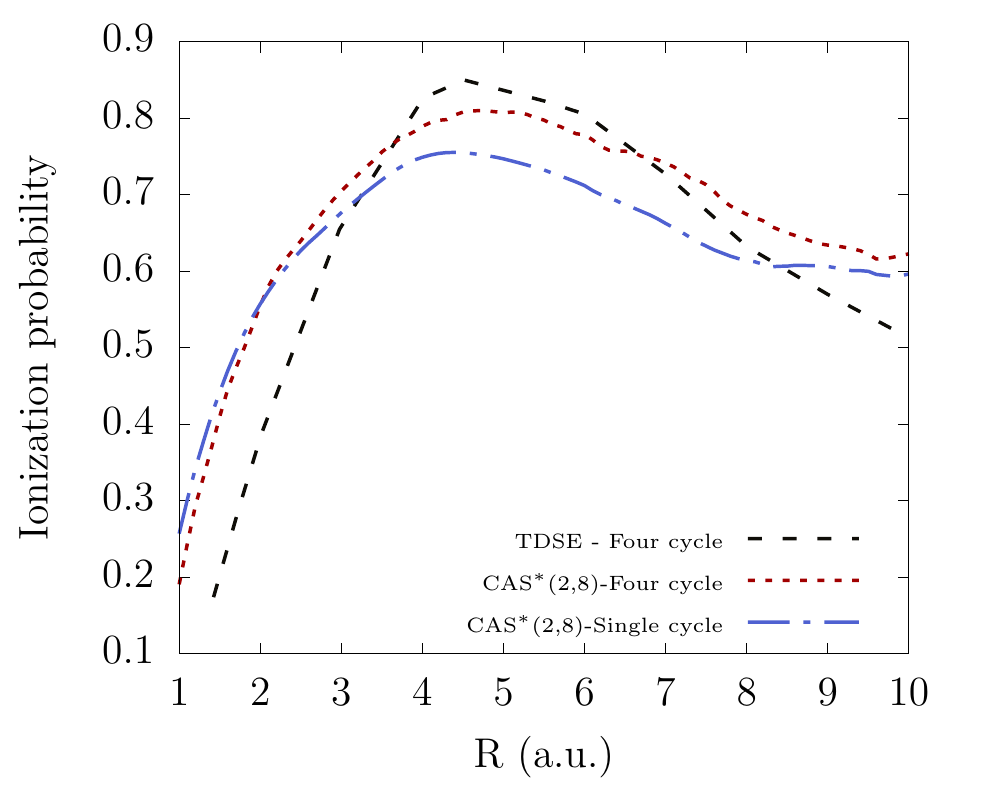}
    \caption{Comparison of the converged CAS $^*(2,8)$ TD-GASCI calculation
	     for the ionization probabilty vs internuclear distance, $R$, 
	     for $\mathrm{H}_2$ with results from full 3D TDSE calculations
	     ~\cite{Dehghanian-PRAR-10}. }
 \label{fig:h2-comp}
\end{figure}

To further test the efficiency of the TD-GASCI method we consider the laser parameters 
from Ref.~\cite{Dehghanian-PRAR-10}. The pulse durations are both 
single- and four-cycle and the peak field strength is 
${F}_0 = 0.053$ ($ 10^{14}$ W/cm$^2$) and $\omega = 0.057$ [Fig.~\ref{fig:e_field}]. 
In Fig.~\ref{fig:h2-comp} we compare the ionization 
probability of $\mathrm{H}_2$ as a function of the internuclear separation 
using the TDSE results provided in Ref.~\cite{Dehghanian-PRAR-10} and the 
converged CAS $^*(2,8)$ model of the TD-GASCI method. There is qualitative 
agreement between both results which further illustrates the capability of the 
TD-GASCI method.

\begin{figure}
    \includegraphics[width=\columnwidth]{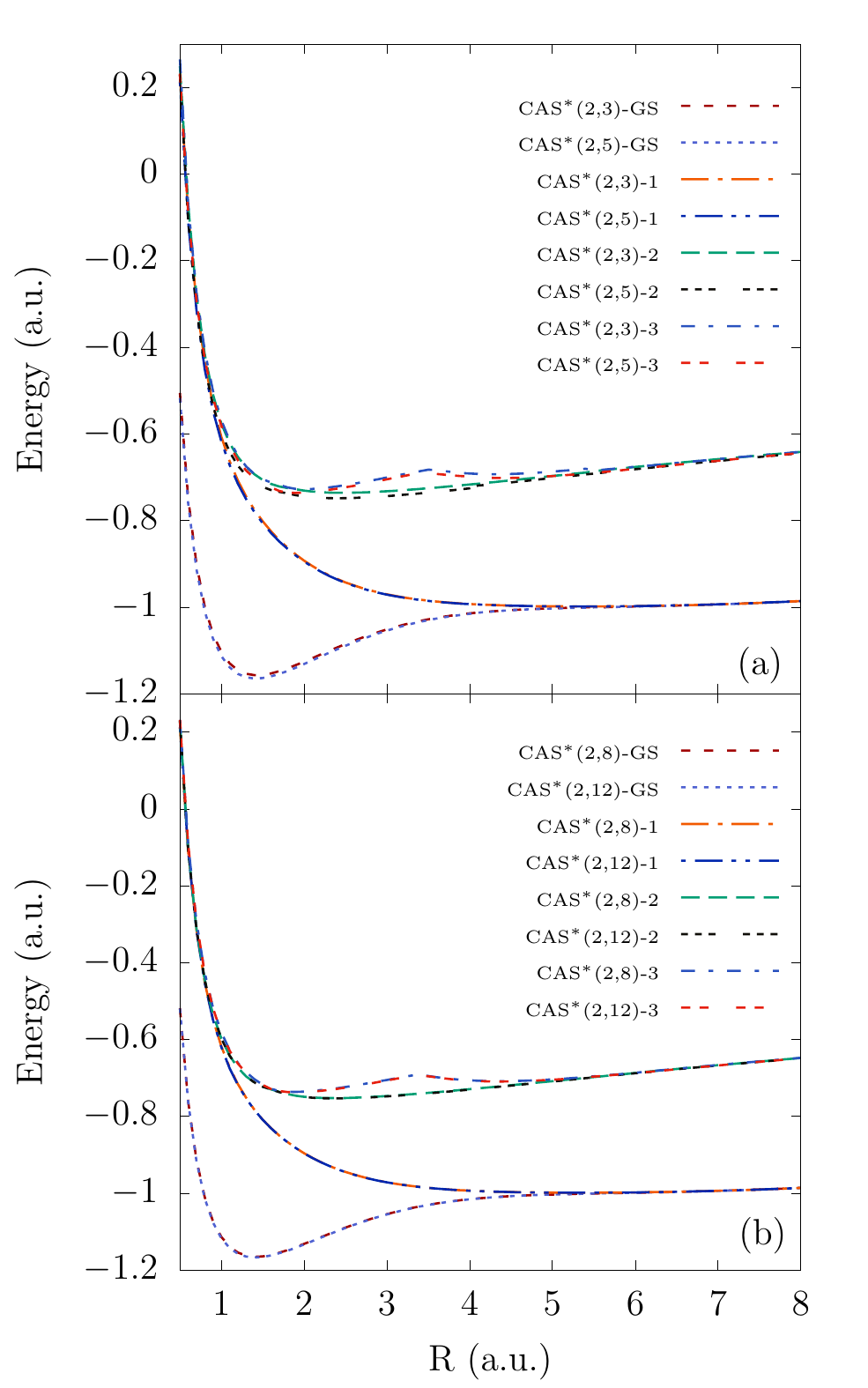}
	\caption{Field-free ground state (GS) and the three lowest-lying 
	excited states of $\mathrm{H}_2$ for (a) CAS$^*(2,3)$ and CAS$^*(2,5)$ and (b) 
	CAS$^*(2,8)$ and CAS$^*(2,12)$. 
	The notation 
		 CAS$^*(2,K)$-$i$ labels the states starting with $i = 1$ for the 
		 first excited state.}
     \label{fig:h2-es-1}
\end{figure}

The electron correlation effect in the EI of  $\mathrm{H}_2$ is 
prominent from the above description. Different CAS-approximations produce the 
correct EI behavior but one can observe that at least eight active orbitals 
are required to properly describe the EI process. On the other hand one needs 
only five active orbitals to have a good description of the ground state. We
 found  that the electronically excited states play
an important role in the EI mechanism. This was discussed in our previous
work with a 1D-$\mathrm{H}_2$-model  molecule, and we observed a similar trend 
in the 3D calculations. The $K$ active orbitals in the GAS partition allow the 
convergence of the electronic excited states for the corresponding 
CAS$^*(2,K)$ model and the same $K$ active orbitals are required for a converged 
EI calculation in the TD-GASCI method. In Fig.~\ref{fig:h2-es-1}(a), we compare
energies of the  lowest four field-free states from CAS$^*(2,3)$ and CAS$^*(2,5)$ 
calculations. These four states are obtained by directly diagonalizing the 
CI-Hamiltonian in a small simulation box which also provides accurate energies. 
As we further increase the number of active orbitals in a GAS partition, we can 
see that the lowest four states converge as shown in Fig.~\ref{fig:h2-es-1}(b). 
Thus we can see that at least eight active orbitals are required to obtain a 
converged result for these low-lying excited states which equals the number of 
states  needed for the convergence of the EI process. This equality
illustrates the role of electronically excited states in the EI mechanism.
One can intuitively interpret that when the strong laser field is applied, the 
low-lying excited states can be involved in a strong coupling with the ground state at some intermediate internuclear separation and this leads to EI.

\subsection{Four-electron $\mathrm{LiH}$ molecule}
\label{lih}
\begin{figure}
 \includegraphics[width=\columnwidth]{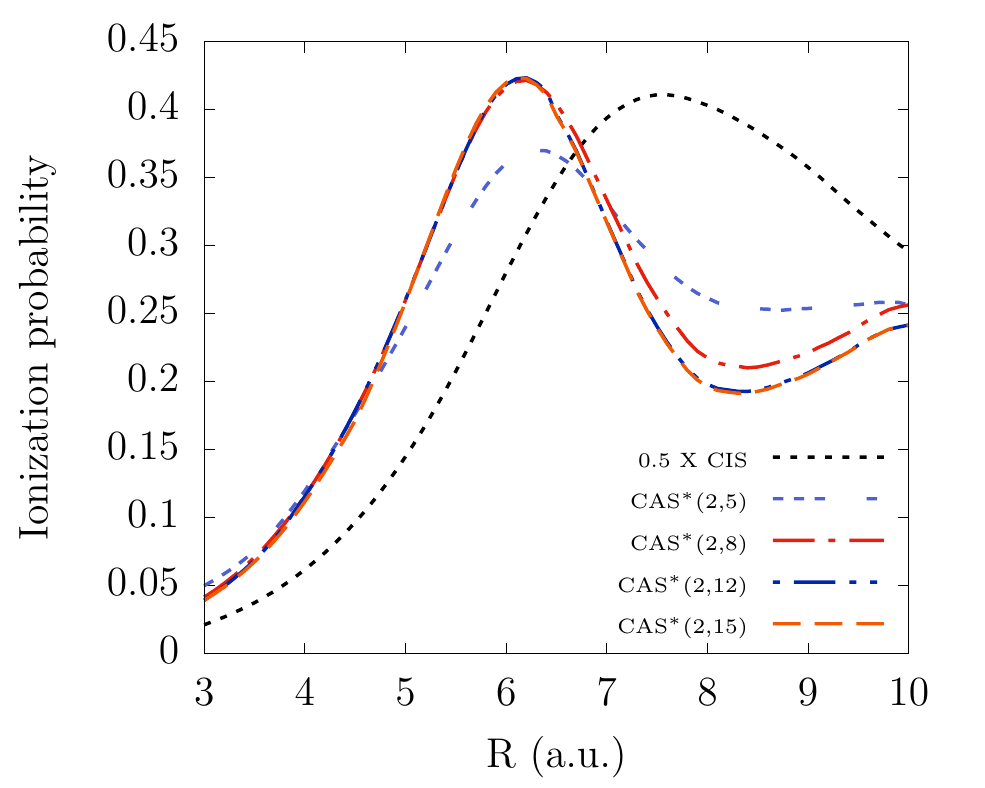}
	\caption{Ionization probability vs internuclear distance 
	     $R$ for $\mathrm{LiH}$ with ${F}_0 = 0.025$ with CIS and different 
	     GAS approximations.}
 \label{fig:lih-ei}
\end{figure}

In this section we present an analysis of EI in the four-electron 
$\mathrm{LiH}$ molecule. It is one of smallest heteronuclear systems which has 
been studied for electron correlation effects in both 1D and 3D calculations 
~\cite{Bauch-PRA-14,Chattopadhyay-PRA-15b,Larsson-PRA-16}. Like in the case of
$\mathrm{H}_2$, orbitals with higher $m$-quantum numbers are 
needed for an accurate description of the electronic correlation and we chose
up to $m = \pm 1$ for the time-dependent calculations. 

We use  ITP to prepare $\mathrm{LiH}$ in its ground state and 
then the laser pulse with a peak field strength of ${F}_0 = 0.025$ 
(2.18$\times 10^{13}$ W/cm$^2$) is applied to ionize the molecule. We compare
the results of EI with CIS and different GAS approximations in 
Fig.~\ref{fig:lih-ei}. Similar to $\mathrm{H}_2$, for $\mathrm{LiH}$ the CIS 
and all GAS approximations predict an EI peak. However, the CIS 
approximation predicts an incorrect EI peak position as well as magnitude
compared to the other more accurate GAS approximations. This results 
further reflects that electron correlation effects should be 
taken into account to explain EI in diatomic molecules. For the CAS$^*(2,5)$ 
scheme, the EI peak is observed at $R = 5.9$. As we further increase the number 
of active orbitals, the peak remains at the same position but the magnitude of 
the ionization probability increases further until convergence is obtained with the CAS$^*(2,8)$ 
scheme. Increasing the number of active orbitals in the GAS partitions 
shows the trend of convergence.  Here we would 
also like to point out that  
the converged peak is shifted from $R = 6.1$ in the 1D calculation~\cite{Chattopadhyay-PRA-15b} to
 $R = 5.9$ in the present 3D case for the same set of 
laser parameters. This  highlights the necessity of using the Coulomb potential instead of 
regularized-Coulomb potential for an accurate description of the EI process.

\begin{figure}
    \includegraphics[width=\columnwidth]{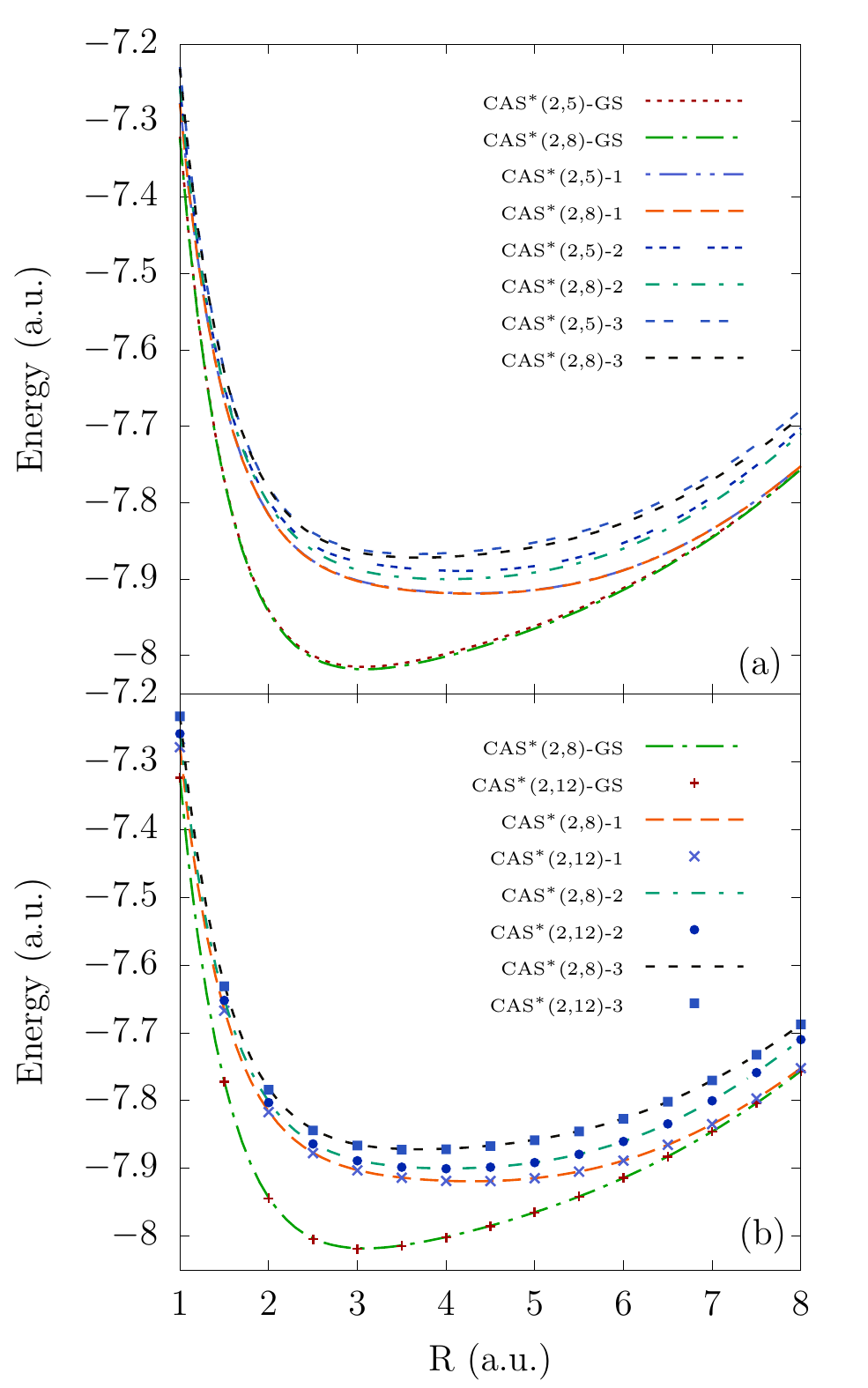}
	\caption{Field-free ground state (GS) and the three lowest-lying
	     excited states of $\mathrm{LiH}$ for (a) CAS$^*(2,5)$ and CAS$^*(2,8)$ and (b) 
	     CAS$^*(2,8)$ and CAS$^*(2,12)$.
             The notation CAS$^*(2,N)$-$i$ labels the states starting 
	     with $i = 1$ for the first excited state.}
     \label{fig:lih-es-1}
\end{figure}

To further study the role of electronic excited states in the EI mechanism of
the $\mathrm{LiH}$ molecule we perform a diagonalization of the 
time-independent CI-Hamiltonian to obtain low-lying excited states. In 
Fig.~\ref{fig:lih-es-1}(a) we show the field-free ground-state and the three 
lowest-lying excited states of $\mathrm{LiH}$ in CAS$^*(2,5)$ and 
CAS$^*(2,8)$ approximations. It is discernible that the CAS$^*(2,5)$ model does 
not produce correct field-free excited states and as we increase the number of 
active orbitals the ground and excited states converge as shown in 
Fig.~\ref{fig:lih-es-1}(b). Here one can see again that we need eight active 
orbitals in the GAS space to obtain a converged result. Note 
that all the energy curves are obtained with two-active electrons, i.e., with 
the CAS$^*(2,K)$ approximations with $K$ the number of 
active spatial orbitals. In the ITP method, we obtain an equilibrium bond-length of 
$R = 3.0$ for $\mathrm{LiH}$, which is very close to the value obtained by quantum chemistry 
calculations~\cite{Helgaker-MEST-14}. One can in principle use four-active 
electron to obtain accurate energy curves in the TD-GASCI method. However, 
as shown in 1D-calculations~\cite{Chattopadhyay-PRA-15b}, the 
four-active electrons situation provide the EI peak at the same position as 
for  two active electrons. Also due to higher computational cost with FE-DVR 
basis in both $\xi$ and $\eta$ -coordinate, we perform the imaginary and real 
time-propagation with two-active electrons.

\subsection{Ten-electron $\mathrm{HF}$ molecule}
\label{hf}
\begin{figure}
 \includegraphics[width=\columnwidth]{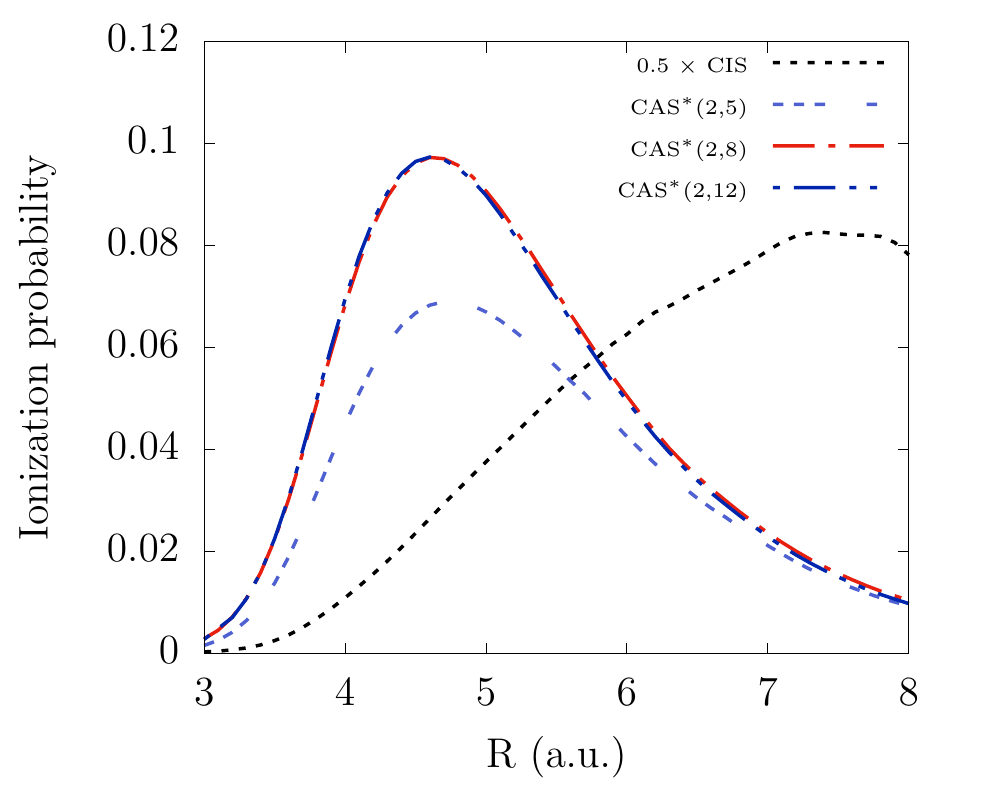}
 \caption{Ionization probability vs internuclear distance $R$ for the $\mathrm{HF}$ molecule
	 with ${F}_0 = 0.05$ with CIS and different CAS$^*(2,K)$ approximations. }
 \label{fig:hf-ei}
\end{figure}

One of the significant advantages of the TD-GASCI method over TDSE is the 
capability of a treatment of atoms and molecules with more than two electrons. 
To verify the universality of the EI process and electron correlation effects in
strong-field ionization of multi-electron molecules, we consider
the $\mathrm{HF}$ molecule. 
\begin{figure}
    \includegraphics[width=\columnwidth]{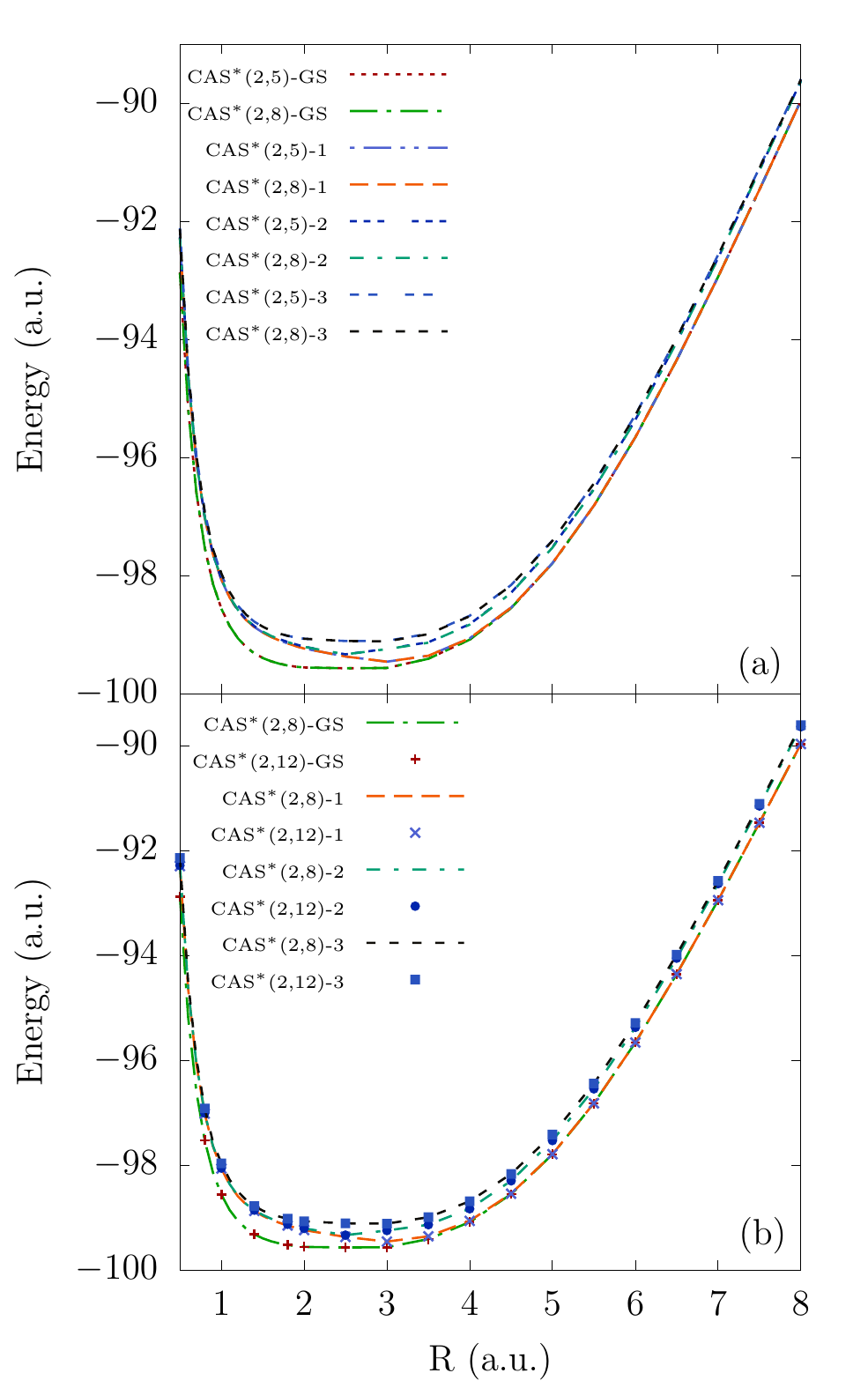}
	\caption{Field-free ground state (GS) and the three lowest-lying excited states of the 
	$\mathrm{HF}$  molecule for (a) CAS$^*(2,5)$ and CAS$^*(2,8)$ and (b) CAS$^*(2,8)$ and CAS$^*(2,12)$.
             The notation CAS$^*(2,N)$-$i$ labels the states starting 
	     with $i = 1$ for the first excited state.}
     \label{fig:hf-es-1}
\end{figure}

Similar to the previous cases, we prepare the $\mathrm{HF}$ molecule in its 
ground state with ITP method. The laser field as shown in 
Fig.~\ref{fig:e_field} (a) is applied with a field strength of ${F}_0 = 0.05$ 
(8.75$\times 10^{14}$ W/cm$^2$) to ionize the molecule. In 
Fig.~\ref{fig:hf-ei}, we show the ionization probability against the internuclear
separation calculations for $\mathrm{HF}$. We compare the CIS approximation 
and different GAS methods. It is clear, like in the previous cases, that the CIS
approximation fails to produce the correct EI peak position and magnitude.
The present result also indicates that as the number of electrons in a
system increases, the correlation effect may become more prominent. Since
the CIS approximation does not take into account the doubly excited Slater
determinants, it fails to incorporate a major part of the electron
correlation. The present calculation further emphasizes the need of 
correlated time-dependent calculations for this kind of  process.
In the GAS schemes, we found that all the methods produce the EI peak at 
$R = 4.6$. The lowest GAS calculation with the CAS$^*(2,5)$ model predicts the 
correct EI peak position. As we increase the number of active orbitals, the 
EI peak converge with the CAS$^*(2,8)$ model as in the previous cases. To check 
the convergence, we increase the number of active orbitals up to twelve as 
shown in Fig.~\ref{fig:hf-ei} and find no significant changes in EI peak. 
Therefore we need eight active orbitals in this case to obtain a converged
results for EI.

To study the role of electronic excited states in the EI of the $\mathrm{HF}$ molecule,
we diagonalize the time-independent CI-Hamiltonian. We find the ground state and three 
lowest-lying excited state energies
shown in Fig.~\ref{fig:hf-es-1}. We note that
in this case the excited states obtained from the CAS$^*(2,5)$ model almost
overlap  with the CAS$^*(2,8)$ model. We find a complete convergence
of the excited states as shown in Fig.~\ref{fig:hf-es-1}(b). Therefore in
this case also the number of active orbitals required to produce converged
excited states produce the converged EI. So we can conclude that in the near
infrared region, along with the ground-state one needs an accurate 
representation of the low-lying excited states to obtain converged results
for the EI process.

\section{Summary and Conclusion}
\label{conc}
The present work highlights the role of electron correlation effects in EI
 of diatomic molecules. The TD-GASCI method based on the 
generalized-active-space concept build electron correlation in a systematic 
way.  First we 
considered $\mathrm{H}_2$  and we found that the SAE and
CIS approximations do not accurately describe EI.
These two approximation even produce some spurious peaks in the EI signal.
The more accurate GAS calculations produce converged ionization probabilities and we found
a very good agreement with previous TDSE calculations~\cite{Dehghanian-PRAR-10}. 
We demonstrated the importance of considering the double excitations in the many-electron
wave function. It is remarkable that electron correlation may reduce the EI probability. 
We highlighted the usefulness 
of the TD-GASCI method which is computationally less expensive than the full 
TDSE treatment. The two-active electron approach was also used to treat 
$\mathrm{LiH}$ and $\mathrm{HF}$ molecules. We demonstrated that EI 
persists in these multielectron molecules and that electron correlation is 
necessary to obtain converged EI results. Also for these two molecules, the CIS 
approximation fails to predict correct EI results. 

The present work is to our knowledge, the 
first that presents ab initio calculations on EI for system larger than $\mathrm{H}_2$. 
The results show that the EI process is universal
and that correlated calculations are needed to accurately describe the process.
We found that  the EI results 
strongly depend on the convergence of the excited states. We conclude that 
to obtain a correct description of EI in near-infrared field, one needs an accurate representation of 
the ground state as well as of low-lying electronically excited states. In the future, we expect that the
TD-GASCI method can be applied to study the importance of
electron correlation effects in EI in mid-infrared regime.

\begin{acknowledgments}
This work was supported by the ERC-StG (Project No. 277767-TDMET), the VKR 
center of excellence, QUSCOPE. The authors thank S.~Bauch,
H.~R.~Larsson and L.~K.~S{\o}rensen for work on the initial implementation of the
TD-GASCI code. The numerical results presented in this work were obtained at the 
Centre for Scientific Computing Aarhus.

\end{acknowledgments}

\appendix
\section{Numerical parameters for discretization}
In this Appendix, we give details on the discretization of the prolate spheroidal coordinates 
and the typical CPU usages for different GAS schemes.
For $\mathrm{H}_2$,  we found that for $\xi$ two
finite elements with 8 and 7 FE-DVR functions in each element with a simulation
box size with $\xi_\text{max}=11$, 10 FE-DVR functions in $\eta$-coordinate
and $m=0, \pm{1}$ is enough to obtain a converged Hartree-Fock energy and 
orbitals. Further increasing the number of FE-DVR functions 
improves the accuracy of the ground-state calculations and in this method one 
can reach the accuracy of different quantum chemistry calculations using a 
significantly higher number of FE-DVR functions within the central region. 
For our study of EI, we are interested in processes involving continuum dynamics and 
this  limits the number of FE-DVR functions that can be used to construct the Hartree-Fock 
orbitals. For the time-dependent part we define the central region up to 11 and 
increase the simulation box size up to 151 in the $\xi$-coordinate. The outer 
region in this case consists of 28 finite elements with seven FE-DVR functions 
in each element. Therefore the full simulation box is of size 151 and it 
contains 181 basis functions in the $\xi$-coordinate. Further we use 10 FE-DVR 
functions in the $\eta$-coordinate and consider $m = 0,\pm{1}$ for the final 
time-dependent simulations. So in total we have 5430 basis functions. We found 
that this ensures converged result for $\mathrm{H}_2$. We emphasize that the 
CAS-calculations performed in the present study are
referred to as correlated CAS-calculations compared to the SAE and CIS
approximations as these two approximations do not include significant
contributions to the dynamic electron correlation arising from double
excitations. 

For all EI calculations, a single-cycle pulse has been used. 
Only for the comparison in Fig.~6,  a four-cycle pulse was used. 
With the  CAS$^*(2,3)$ model, which is the smallest CAS 
calculation performed in the present work, and using the feature of the Intel 
MKL-library for sparse matrix-vector multiplication in a Intel Ivy-bridge 
processor with 20 cores at 2.8 GHz speed it takes  3 hours, 14 minutes to 
complete a simulation. With the CAS$^*(2,12)$ model it takes 13 hours 
and 3 minutes to finish. For the four-electron $\mathrm{LiH}$ we found that two 
finite-elements with 14 and 17 FE-DVR functions with a simulation box size of
14 is sufficient to produce converged results in the central region. Similar to
$\mathrm{H}_2$, we use 10 FE-DVR functions for the $\eta$-coordinate and we
consider $m = 0, \pm{1}$. For the time-dependent calculations we choose a
simulation box with $\xi_\text{max}= 150$. The total simulation box in 
this case has 182 FE-DVR functions for the $\xi$-coordinate and 10 
FE-DVR functions for the $\eta$-coordinate and $m = 0, \pm{1}$. For the
full simulation box we thus have  5460 basis functions and observe 
converged results for all CAS-calculations. For this molecule, a
time-dependent calculation with the CAS$^*(2,5)$ model with the same computational
configuration takes 21 hours and 7 minutes to finish. The largest 
CAS$^*(2,15)$-scheme has taken 11 days and 1 hour and 16 minutes to obtain a 
converge result. In case of the $\mathrm{HF}$ molecule, we found that two finite-elements and 
20 and 7 FE-DVR functions in each element for the $\xi$-coordinate and 12 FE-DVR 
functions in the $\eta$-coordinate and $m=0,\pm{1}$ ensure a converged 
Hartree-Fock energy and orbitals. For the time-dependent calculations we use a 
simulation box of size 101 and 133 FE-DVR functions in the $\xi$-coordinate 
and 12 FE-DVR functions in the $\eta$-coordinate and $m=0,\pm{1}$. The total number 
of basis functions in this case is 4788. A time-dependent 
calculation with the smallest CAS$^*(2,5)$ model takes 5 days, 10 hours and 58 
minutes to complete. For the largest CAS$^*(2,12)$-scheme the time-dependent 
calculation take 21 days, 4 hours and 16 minutes to finish. The 
computational cost therefore restricts us to consider the four-active electron 
situation and in Ref.~\cite{Larsson-PRA-16}, it was found that for 
$\mathrm{LiH}$, the CAS$^*(2,8)$ model produces the same ionization probability as the
CAS$^*(4,4)$ model. Therefore all our calculations are performed with two-active 
electrons.

%

\end{document}